# Title: Spin-Layer Locking Effects in Optical Orientation of Exciton Spin in Bilayer WSe$_2$

**Authors:** Aaron M. Jones[1], Hongyi Yu[2], Jason S. Ross[3], Philip Klement[1,4], Nirmal J. Ghimire[5,6], Jiaqiang Yan[6,7], David G. Mandrus[5-7], Wang Yao[2], Xiaodong Xu[1,3*]

**Affiliations:**

[1]Department of Physics, University of Washington, Seattle, Washington 98195, USA
[2]Department of Physics and Center of Theoretical and Computational Physics, University of Hong Kong, Hong Kong, China
[3]Department of Materials Science and Engineering, University of Washington, Seattle, Washington 98195, USA
[4]Department of Physics, Justus-Liebig-University, 35392 Giessen, Germany
[5]Department of Physics and Astronomy, University of Tennessee, Knoxville, Tennessee 37996, USA
[6]Materials Science and Technology Division, Oak Ridge National Laboratory, Oak Ridge, Tennessee 37831, USA
[7]Department of Materials Science and Engineering, University of Tennessee, Knoxville, Tennessee 37996, USA

*Correspondence to: xuxd@uw.edu

**Abstract:** Coupling degrees of freedom of distinct nature plays a critical role in numerous physical phenomena[1–10]. The recent emergence of layered materials[11–13] provides a laboratory for studying the interplay between internal quantum degrees of freedom of electrons[14,15]. Here, we report experimental signatures of new coupling phenomena connecting real spin with layer pseudospins in bilayer WSe$_2$. In polarization-resolved photoluminescence measurements, we observe large spin orientation of neutral and charged excitons generated by both circularly and linearly polarized light, with a splitting of the trion spectrum into a doublet at large vertical electrical field. These observations can be explained by locking of spin and layer pseudospin in a given valley[15]. Because up and down spin states are localized in opposite layers, spin relaxation is substantially suppressed, while the doublet emerges as a manifestation of electrically induced spin splitting resulting from the interlayer bias. The observed distinctive behavior of the trion doublet under circularly and linearly polarized light excitation further provides spectroscopic evidence of interlayer and intralayer trion species, a promising step toward optical manipulation in van der Waals heterostructures[16] through the control of interlayer excitons.

**Main Text**

Exploring the consequences of the interplay between distinct quantum degrees of freedom has been an active theme in modern physics. A salient example is spin-orbit coupling (SOC), which is essential in renowned condensed matter phenomena such as the spin Hall effect[1,2], topological insulators[3,4] and Majorana fermions[5,6]; in cold atom physics in the search for new condensate structures[7]; and in technological applications such as magnetoelectric coupling in multiferroics[8] as well as optical and electrical control of spins for spintronics[9,10]. All these phenomena arise from the coupling of the motional degree of freedom of a particle with its real spin.

A pseudospin describes another discrete internal degree of freedom of electrons, and in most systems has an orbital origin and can therefore couple to the real spin by SOC as well. An excellent example is afforded by monolayer transition metal dichalcogenides (TMDCs), which have attracted a significant amount of interest recently[12,13,17–21]. The inversion symmetry breaking allows for an effective coupling between the real spin and valley pseudospin[14] (the latter indexes the degenerate extrema of the electron energy dispersion in momentum space). In the presence of mirror and time-reversal symmetry, SOC can be manifested as an out-of-plane spin splitting with a valley-dependent sign (c.f. Fig. 1a).

Bilayer 2D materials (e.g. bilayer graphene[22–26] and bilayer TMDCs[15]) possess another distinct degree of freedom known as the layer pseudospin. An electronic state localized to the upper or lower layer can be labeled with pseudospin up or down, respectively, which corresponds to electrical polarization. In a layered material with spin-valley coupling and AB stacking, such as bilayer TMDCs, both spin and valley are coupled to layer pseudospin[15]. As shown in Fig. 1a, since the lower layer is a 180º in-plane rotation of the upper layer, the out-of-plane spin splitting has a sign which depends on both valley and layer pseudospins. Interlayer hopping thus has an energy cost equal to twice the SOC strength $\lambda$. When $2\lambda$ is larger than the hopping amplitude $t_\perp$, a carrier is localized in either the upper or lower layer depending on its valley and spin state. In other words, in a given valley, the spin configuration is locked to the layer index. This is schematically illustrated in Figs. 1b and 1c. This spin-layer locking permits electrical manipulation of spins via gate control of layer polarization, which may lead to novel magnetoelectric effects and quantum logic[15].

In this paper we report experimental signatures of coupling between this layer pseudospin and the spin and valley degrees of freedom in bilayers of WSe$_2$ (bi-WSe$_2$). In contrast to MoS$_2$[19], the high quality of WSe$_2$, in addition to much larger SOC, provides an excellent system for observing spin-layer locking. Although bi-WSe$_2$ is an indirect bandgap semiconductor, the near degeneracy between indirect and direct transitions permits us to efficiently monitor direct-gap photoluminescence (PL) from the K valleys[12,13,27]. Electrons in the conduction bands near the K valleys exhibit a spin splitting $2\lambda \sim 30-40$ meV, which is two orders of magnitude larger than the interlayer hopping amplitude $t_\perp$ at the ±K points (Supplementary Section S1). For holes, $2\lambda \sim 450$ meV and $2\lambda/t_\perp \sim 7$ [15]. Large $2\lambda$ to $t_\perp$ ratios ensure that interlayer hopping for both electrons and holes is suppressed, achieving a spin-layer locking effect. In our experiment, bi-WSe$_2$ was obtained through mechanical exfoliation[11] of WSe$_2$ crystals onto 300 nm SiO$_2$ on a Si substrate, followed by patterning into field effect transistors using standard electron beam lithography. All studies were performed at a temperature of 30 K with 1.88 eV laser excitation and 1.5 µm spot size, unless noted otherwise.

We first identify the exciton states in bi-WSe$_2$ through gate-dependent PL measurements[28,29]. Figure 2a shows the PL intensity from direct gap exciton emission as a function of back gate voltage ($V_g$) and photon energy. By comparing the gate dependent patterns and emission energies of monolayer (Supplementary Section S2)[29] and bilayer WSe$_2$, we can identify the weak feature near 1.74 eV and $V_g = 0$ as neutral exciton (X$^o$) emission, while the peak near 1.71 eV at positive or negative $V_g$ corresponds to negative (X$^-$) or positive (X$^+$) trions, respectively. The PL peak near 1.63 eV arises from impurity-bound excitons. Figure 2c shows PL spectra at three selected values of $V_g$. The peaks here coincide with the lowest energy absorption feature shown in Fig. 2b, the gate dependent differential reflectivity obtained by white light reflection (full spectrum in Supplementary Section S3).

Next we show that large and electrically tunable exciton spin polarization can be generated in bi-WSe$_2$ by optical pumping. Figure 3a shows the polarization-resolved PL under $\sigma^+$ excitation for $\sigma^+$ (black curve) and $\sigma^-$ (red curve) detection, at selected $V_g$. The degree of polarization is large for X$^o$, X$^+$, and X$^-$ at all voltages, demonstrating the generation of large exciton spin polarizations. Interestingly, for $V_g$ above 60 V, a doublet structure emerges in the X$^-$ spectrum for $\sigma^+$ (co-polarized) detection, with an increasing separation between peaks at higher $V_g$. We label the peaks I and II as shown in the bottom right of Fig. 3a. Note that the position of peak II coincides with the single peak present for σ⁻ (cross-polarized) detection. The doublet can be fitted by a dual Lorentzian line shape (blue lines).

We define the degree of circular polarization as $\eta_\sigma = \frac{PL(\sigma^+) - PL(\sigma^-)}{PL(\sigma^+) + PL(\sigma^-)}$ where $PL(\sigma^\pm)$ is the detected PL with $\sigma^\pm$ polarization. The extracted $\eta_\sigma$ of the trion peak as a function of $V_g$ is shown in Fig. 3b. For $V_g$ below 60 V, the doublet separation is not resolved and $\eta_\sigma$ is obtained at the peak position without fitting. The inset shows a zoomed-in view of $\eta_\sigma$ centered at 0 V, where an enhancement of an already large $\eta_\sigma$ for increasing $|V_g|$ is observed. The doublet emerges for $V_g$ above 60 V with differing degrees of exciton spin orientation for the two peaks. For simplicity, we plot just the $\sigma^+$ branch of $\eta_\sigma$, with the black and blue dots indicating the polarization of trion peaks I and II, respectively, obtained from the peak fittings (see Supplementary Section S4 for fitting parameters). Figure 3c shows the peak splitting as a function $V_g$ under $\sigma^+$ polarized excitation (black squares).

The gate dependent PL in Fig. 2a shows the crossover between X⁺ and X⁻ near $V_g = 0$, which demonstrates that the sample is nearly intrinsic, without substantial external doping from substrate effects or impurities. The inset to Fig. 3b shows that the polarization minimum is near $V_g = 0$. We also performed second harmonic generation (SHG) measurements (Supplementary Section S5), which showed more than two orders of magnitude suppression of SHG in bi-$WSe_2$ compared to monolayer[30]. These results imply the presence of inversion symmetry in unbiased bi-$WSe_2$. Therefore, it is not possible that the circularly polarized PL near $V_g = 0$ stems from valley polarization, as demonstrated in single layer materials[17–19], which requires explicit inversion symmetry breaking[14]. Rather, it originates from exciton spin polarization, a consequence of the spin optical selection rules present for both inversion-symmetric and asymmetric bilayers with large spin-orbit coupling[15].

We attribute the large exciton spin polarization, together with the emergence of an X⁻ doublet at high electric field, to the spin-layer locking effect[15], which leads to an enhanced spin lifetime and electrically-induced spin splitting. Figure 3d shows the energy level diagram of AB stacked bi-$WSe_2$ without an applied electric field. Single and double arrows denote the spin configurations of electrons and holes, respectively. Since there are many possible trion configurations, we only show electron-hole pairs which emit significant PL. Under $\sigma^+$ excitation, the transition involving spin up hole states is excited in both +K and -K valleys with equal strength and no valley polarization is generated. Because of the spin-layer locking (c.f. Figs. 1b-c), intra-valley spin flips are suppressed as the spin up and down states are localized in opposite layers. Accordingly, the spin relaxation time can be long compared to the exciton lifetime, leading to large exciton spin orientation[31].

A perpendicular electric field creates a potential difference between upper and lower layers, which lifts the energy degeneracy between spin up and down states localized in opposite layers for a given valley. The result is an electrically induced spin Zeeman splitting. The spin splitting for electrons, $\Delta_c$, is larger by a few percent than that for holes, $\Delta_v$, due to the larger $2\lambda/t_\perp$ ratio (Supplementary Section S6). This difference leads to two distinct emission frequencies, i.e. a higher frequency $\omega_1$ for electron-hole recombination in the upper layer and a lower emission frequency $\omega_2$ for the lower layer (see Fig. 3e). When $\omega_1 - \omega_2$ becomes larger than the spectral line width, the trion peak splits into a doublet as shown in Fig. 3a.

With this doublet resolved at large gate values, we observe differing degrees of polarization for the two trion peaks, which further corroborates spin-layer locking. Specifically, the polarization of peak I centered at the higher frequency $\omega_1$ is larger than that of peak II centered at $\omega_2$. As the spectra in Fig. 3a show, for $V_g > 50$ V, the weak PL feature acquired under $\sigma^+$ excitation with $\sigma^-$ detection (red curves), is always centered at $\omega_2$. Figure 3e depicts the mechanism for the reduced polarization of peak II compared to peak I. Considering $\sigma^+$ excitation, $\sigma^+$ polarized PL at $\omega_1$ and $\omega_2$ comes from electron-hole recombination in the upper layer in the -K valley and the lower layer in the +K valley, respectively. Via a spin-flip and dissipation of the energy $\Delta_c$ to the environment, the photo-excited electron in the upper layer of the -K valley can relax to the lower layer within the same valley, which leads to $\sigma^-$ PL at energy $\omega_2$, corresponding to peak II (Fig. S7). In contrast, $\sigma^-$ PL at energy $\omega_1$ requires absorption of the energy $\Delta_c$ to flip the photo-excited electron spin from the lower to the upper layer in the +K valley. Such spin-flip processes are strongly suppressed as $\Delta_c \gg k_B T$ under the applied field necessary to resolve the doublet. Therefore, PL at $\omega_1$ exhibits larger $\eta_\sigma$ than at $\omega_2$.

In light of the clear asymmetry of trion peak splitting with applied gate, we note that the energy difference between $\omega_1$ and $\omega_2$ also has a contribution from the different binding energies for each trion configuration. We consider the lowest energy configurations of optically active trions only, where the extra electron or hole is in the lowest energy band (c.f. Fig. 3e and Fig. S8). For the X⁻ configuration that emits at $\omega_1$, the electron-hole pair is in the upper layer and the excess electron is in the lower layer (interlayer trion). Conversely, all three particles are localized in the lower layer for X⁻ at $\omega_2$ (intralayer trion). The larger wave function overlap in the latter case leads to larger trion binding energies for peak II than peak I. This effect enhances the X⁻ energy splitting, i.e. $\omega_1 - \omega_2 = (\Delta_c - \Delta_v) + |\delta E_B|$, where $\delta E_B$ denotes the binding energy difference between interlayer and intralayer trions. Similar analysis for the X⁺ trion shows that $\omega_1 - \omega_2 = (\Delta_c - \Delta_v) - |\delta E_B|$ (Supplementary Section S8). From the measured trion binding energy of ~30 meV, we estimate $\delta E_B$ to be several meV. Thus the binding energy difference may cancel the electric field induced splitting and lead to the negligible splitting of X⁺ we observe.

We also examined the degree of linear polarization of the PL under linearly polarized excitation, which provides clear evidence for the observation of intralayer and interlayer trions, in addition to further revealing the electrically tunable optical orientation of in-plane excitonic spin and confirming the spin-layer locking effect. Fig. 4a shows polarization-resolved PL spectra at selected $V_g$ under vertically polarized excitation and for vertically ($\updownarrow$) and horizontally ($\leftrightarrow$) polarized detection. In contrast to the observation of linearly polarized PL for only neutral excitons in monolayer WSe$_2$[29], we observe strong linear polarization for both neutral and charged excitonic states in bi-WSe$_2$ (see graph for $V_g = 0$, Fig. 4a). Measuring emitted PL for arbitrary incident polarization angles (green arrows) shows that the trion PL polarization direction always coincides with that of the incident light (Fig. 4b). This demonstrates insensitivity of the PL polarization to sample orientation and thus rules out crystal anisotropy as an explanation.

Defining the degree of linear polarization as $\eta_\updownarrow = \frac{PL(\updownarrow) - PL(\leftrightarrow)}{PL(\updownarrow) + PL(\leftrightarrow)}$ where $PL(\updownarrow)$ and $PL(\leftrightarrow)$ indicate co-linear and perpendicular polarization detection, respectively, we extract the degree of polarization $\eta_\updownarrow$ of trions as a function of gate, shown in Fig. 4c. We find that $\eta_\updownarrow$ levels out to 0.25 for $V_g < 0$, and increases monotonically for $V_g > 0$. For $V_g > 60$ V, the doublet emerges and the peak splitting matches that obtained under circularly polarized excitation (red squares Fig. 3c). The polarization is calculated for each doublet peak after fitting. Interestingly, $\eta_\updownarrow$ of peak I continues to increase monotonically whereas for peak II it hovers near zero. This can easily be seen by polarization-resolved PL under vertically polarized excitation at $V_g = 150$ V (Fig. 4a). We obtain the vertically polarized component at $\omega_2$ (blue line) by fitting the data taken with vertically polarized detection (black curve). The extracted vertically polarized PL at $\omega_2$ matches the measured horizontally polarized component (red dots), indicating negligible linear polarization of peak II. Applying the same procedure at other $V_g$, we determine that linearly polarized emission only comes from peak I.

As demonstrated in monolayer $WSe_2$[29], isotropic linear polarization can arise from the optical generation of a coherent superposition of excitonic states in the K and -K valleys, which must have identical emission energies and final electronic states upon electron-hole recombination. For $X^+$ in bi-$WSe_2$, this is only allowed for the interlayer trion configuration (Supplementary Section S9). For $X^-$, the lower panel of Fig. 4d shows the ground state configuration for emission at $\omega_2$, which is an intralayer trion with all three particles localized in the same lower layer. Because the exchange interaction with the excess electron destroys the inter-valley coherence of the electron-hole pair, no linear polarization is observed at $\omega_2$. Intralayer trion configurations are basically the same as those in monolayer $WSe_2$, and our observation here is in agreement with the absence of linearly polarized $X^-$ PL in monolayers[29]. In contrast, for interlayer $X^-$ emission at $\omega_1$ (upper part of Fig. 4d), the excess electron is in the layer opposite the electron-hole pair and the exchange interaction as a dephasing mechanism between $\sigma^+$ and $\sigma^-$ emission is largely suppressed. As a result, emission at $\omega_1$ is linearly polarized and signals the first observation of an interlayer excitonic state.

In summary, our experiments reveal new coupling of distinct quantum degrees of freedom unique to bilayer TMDCs, i.e. strong coupling between spin and layer pseudospin, which leads to suppression of spin relaxation. The observed peak splitting with applied electric field implies spectral discrimination of layer index, and suggests an electrical means of controlling spin states through spin-layer locking. Additionally, we establish for the first time the creation of a coherent superposition of distinct valley configurations of interlayer trions, where the comprising photo-excited electron-hole pairs and third carrier must be localized in opposite layers so that exchange-induced valley dephasing is suppressed. The demonstrated robust optical selection rules and strong spin-valley-layer coupling readily lend bi-$WSe_2$ to novel spintronic applications.

**Acknowledgements:** The authors wish to acknowledge Guibin Liu and Xianxin Wu for helpful information on the bilayer band structure, and David Cobden for proof-reading. This work is mainly supported by US DoE, BES, Division of Materials Sciences and Engineering (DE-SC0008145). A.M.J. partially supported by NSF graduate fellowship (DGE-0718124). H.Y. and W.Y. were supported by the Research Grant Council (HKU705513P) of the government of Hong Kong, and the Croucher Foundation under the Croucher Innovation Award. N.G., J.Y., and D.M. were supported by US DoE, BES, Materials Sciences and Engineering Division. Device fabrication completed at the University of Washington Microfabrication Facility and NSF-funded Nanotech User Facility. Second harmonic generation is done at Garvey Imaging Core of the Institute for Stem Cell and Regenerative Medicine.

**Author Contributions**

All authors made critical contributions to the work.

**Competing Financial Interests**

The authors declare no competing financial interests.

# Figures

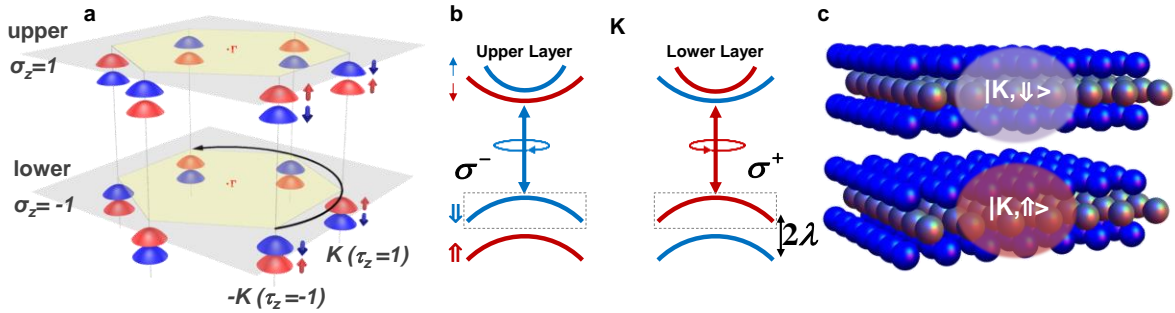

**Figure 1| Coupled spin, valley and layer degrees of freedom in bilayer WSe$_2$. a**, AB stacking order in bilayer TMDCs corresponds to 180º rotation of the lattice between layers, leading to an effective layer pseudospin $\sigma_z$. **b**, Cartoon depicting excitation/emission processes in the K valley of bilayer WSe$_2$. Spin configuration is indicated by ⇑ (↑) for holes (electrons). The same for -K valley is obtained by time reversal. **c**, Depiction of spin down (up) hole states localized in the upper (lower) layer in K valley.

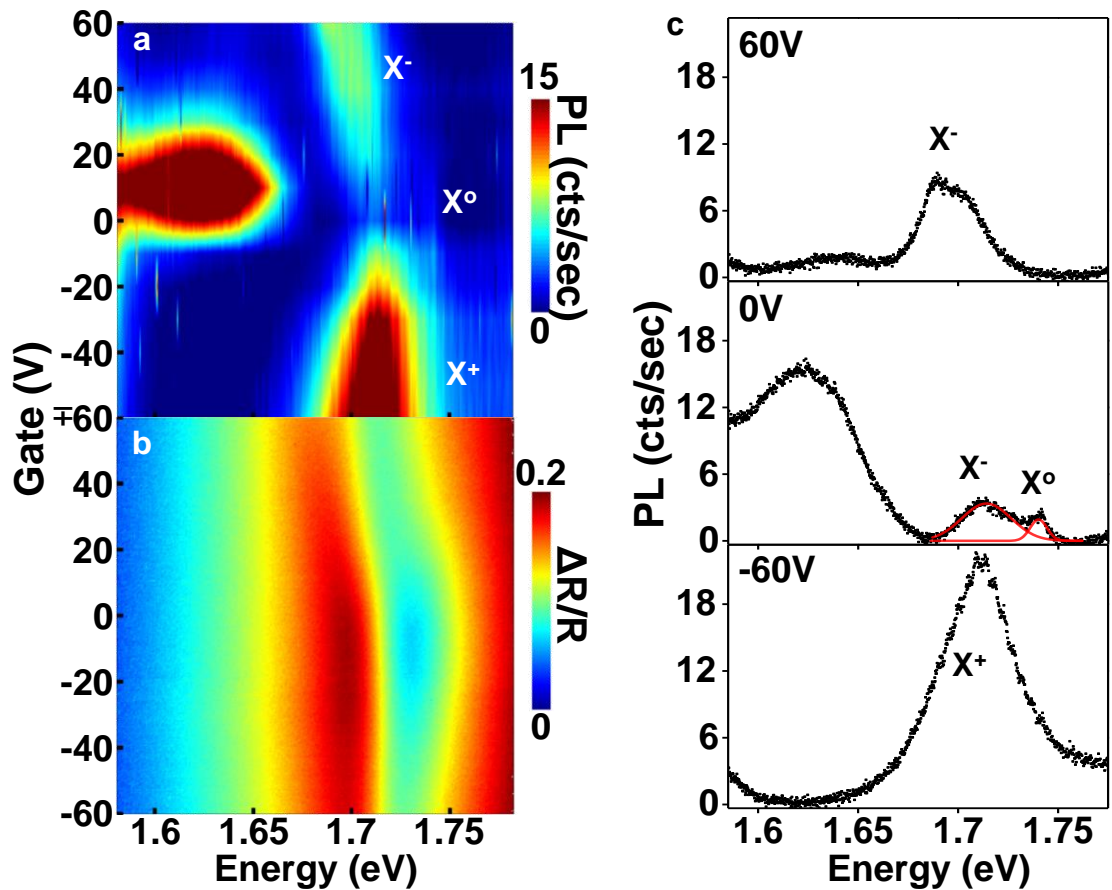

**Figure 2| Photoluminescence and differential reflectivity vs. gate. a**, Photoluminescence intensity as a function of gate voltage and photon energy with labeled neutral exciton ($X^o$), and negative ($X^-$) and positive ($X^+$) trion emission. **b**, Corresponding differential reflectance showing relationship of $X^-$/$X^+$ emission to the absorption feature. **c**, Photoluminescence spectra extracted at +60 V, 0 V and -60 V, showing excitonic peaks $X^-$, $X^o$ and $X^+$, respectively. Red lines in 0 V spectrum show Lorentzian peak fits.

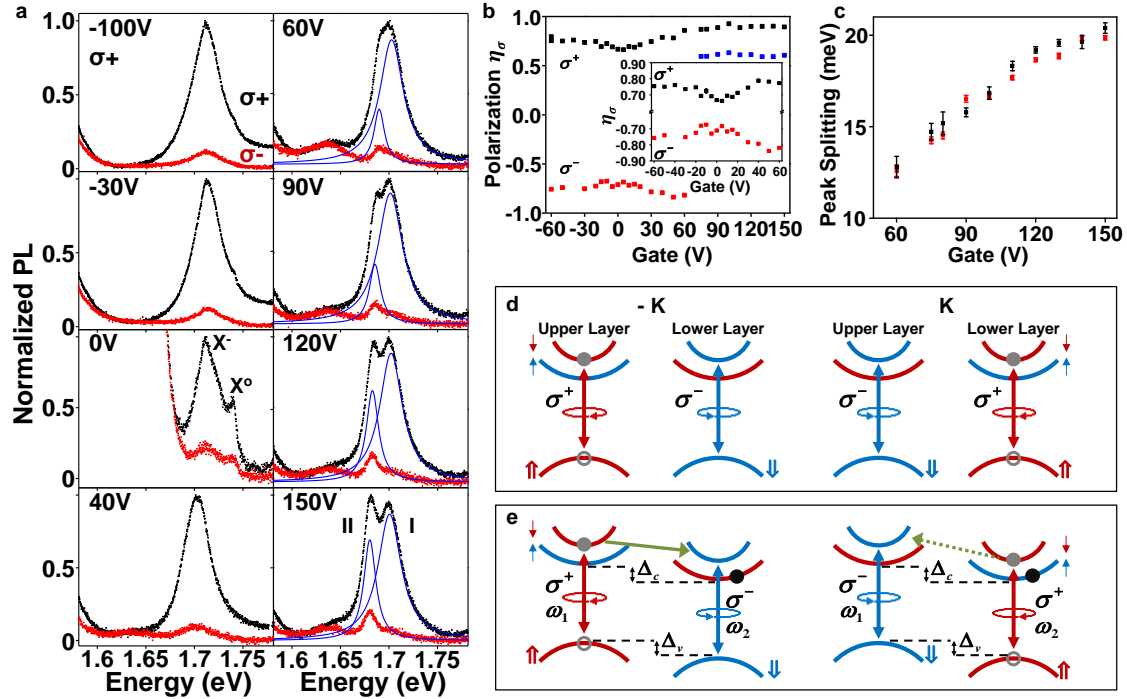

**Figure 3| Optical orientation of spin and gate-induced peak splitting in trions. a**, Normalized photoluminescence spectra vs. photon energy at selected gate voltages for $\sigma^+$ polarized excitation and $\sigma^+$ (black curve) and $\sigma^-$ (red curve) detection. Blue lines in plots for $V_g > 50$ V show Lorentzian peak fits of $\sigma^+$ detection data. **b**, Trion PL circular polarization, $\eta_\sigma$, as a function of gate voltage, where black (red) data indicate $\sigma^+$ ($\sigma^-$) excitation. For $V_g > 50$ V, the trion peak splits and black (blue) data correspond to $\eta_\sigma$ of peak I (II) under $\sigma^+$ excitation. Inset: Zoom-in showing enhancement of $\eta_\sigma$ with applied gate voltage for both $\sigma^+$ (black) and $\sigma^-$ (red) excitation. **c**, Trion peak splitting as a function of applied gate voltage for circularly (black) and linearly (red) polarized excitation. **d**, Schematic depicting the formation of excitons in both K and -K valleys under $\sigma^+$ excitation in unbiased bilayer WSe$_2$, resulting in no net valley polarization. Hollow (solid) circles denote holes (electrons). Gray circles denote photo-excited electron-hole pairs. **e**, Schematic of electric field-induced band shifts and electron spin relaxation pathways (green arrows). Emission from the upper and lower layers is at $\omega_1$ and $\omega_2$, respectively, whose splitting originates in the difference between the conduction ($\Delta_c$) and valence ($\Delta_v$) band energy shifts with gate electric field.

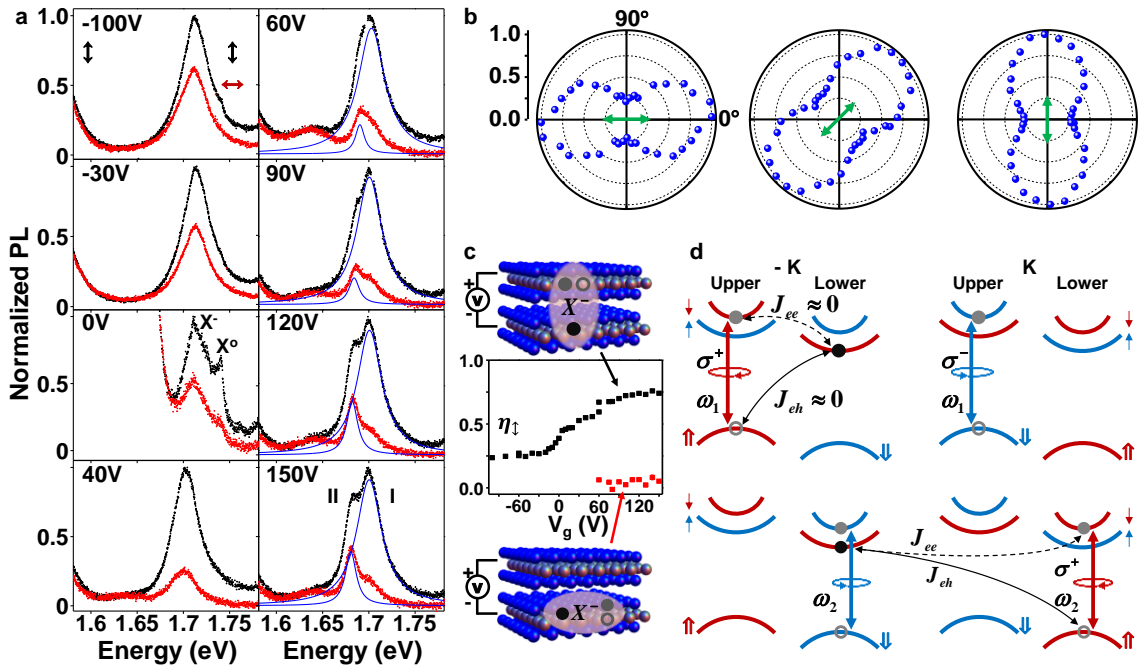

**Figure 4| Linearly polarized excitation of interlayer and intralayer trions. a**, Normalized photoluminescence spectra vs. photon energy at selected gate voltages for vertically ($\updownarrow$) polarized excitation and vertical ($\updownarrow$, black) and horizontal ($\leftrightarrow$, red) detection. Blue lines in plots for $V_g > 50$ V show Lorentzian peak fits of data for vertically polarized detection. **b**, Polar plots showing normalized magnitude of trion peak height as a function of detection angle, at $V_g = 0$. Green arrow indicates incident polarization direction. **c**, Trion linear polarization, $\eta_\updownarrow$, as a function of gate voltage for vertically polarized excitation (center), with corresponding depictions of interlayer (top) and intralayer (bottom) trions. For $V_g > 50$ V, the trion peak splits. Black and red data correspond to $\eta_\updownarrow$ of peak I (interlayer trions) and peak II (intralayer trions), respectively. **d**, Cartoons depicting a superposition of valley trion configurations giving rise to trion peaks I (top) and II (bottom). Due to negligible exchange interactions ($J_{eh}$, $J_{ee}$), surviving inter-valley coherence for interlayer trion configurations leads to large linear polarization at $\omega_1$ (top), while the presence of finite exchange interactions for intralayer configurations eliminates linear polarization at $\omega_2$ (bottom).

# Spin-Layer Locking Effects in Optical Orientation of Exciton Spin in Bilayer WSe$_2$

Aaron M. Jones, Hongyi Yu, Jason S. Ross, Philip Klement, Nirmal. J. Ghimire,

Jiaqiang Yan, David G. Mandrus, Wang Yao, Xiaodong Xu

## S1. Interlayer tunneling in the K valleys of AB stacked bilayers

We first analyze the band edge wavefunctions in the K valleys of two decoupled monolayers with AB stacking order. Here, the conduction band edges originate predominantly from the transition metal $d_{z^2}$ orbital (with magnetic quantum number $m = 0$) [S1], and the Bloch function can be written as $\psi_{c\mathbf{K}}^{u,l} = \sum_n e^{i\mathbf{K}\cdot\mathbf{r}_n} d_{m=0}^{u,l}(\mathbf{r}_n)$. The vector $\mathbf{r}_n$ is the location of the $n$-th metal atom, and superscript $u$ ($l$) denotes the upper (lower) layer. The valence band edges in the upper (lower) layer originate predominantly from the transition metal $d_{x^2-y^2} - id_{xy}$ ($d_{x^2-y^2} + id_{xy}$) orbital [S1] with magnetic quantum number $m = -2$ ($m = +2$). The corresponding Bloch function at $\mathbf{K}$ is then $\psi_{v\mathbf{K}}^u = \sum_n e^{i\mathbf{K}\cdot\mathbf{r}_n} d_{m=-2}^u(\mathbf{r}_n)$ ($\psi_{v\mathbf{K}}^l = \sum_n e^{i\mathbf{K}\cdot\mathbf{r}_n} d_{m=+2}^l(\mathbf{r}_n)$), and those at the $-\mathbf{K}$ point are just their time reversal.

Now we consider nearest neighbor interlayer hopping for the metal atoms, as shown in Fig. S1. The three pairs of interlayer hopping shown by the arrows in Fig. S1 are related by $2\pi/3$ rotations because of the crystal's three-fold rotational symmetry. The center upper layer atom has position $\mathbf{r}=0$ while the positions of the three lower layer atoms are denoted as $\mathbf{r}_1$, $\mathbf{r}_2$ and $\mathbf{r}_3$. The hopping amplitude is proportional to the wave function overlap, given by $\sum_{n=1,2,3} e^{i\mathbf{K}\cdot\mathbf{r}_n} \langle d_{m=0}^l(\mathbf{r}_n) | d_{m=0}^u(0) \rangle$ for the conduction band, and $\sum_{n=1,2,3} e^{i\mathbf{K}\cdot\mathbf{r}_n} \langle d_{m=+2}^l(\mathbf{r}_n) | d_{m=-2}^u(0) \rangle$ for the valence band.

Writing $\mathbf{K} = \frac{4\pi}{3a}\hat{\mathbf{x}}$, then $e^{i\mathbf{K}\cdot\mathbf{r}_1} = 1$, $e^{i\mathbf{K}\cdot\mathbf{r}_2} = e^{-i2\pi/3}$ and $e^{i\mathbf{K}\cdot\mathbf{r}_3} = e^{i2\pi/3}$. $\langle d_{m'}^l(\mathbf{r}_2) | d_m^u(0) \rangle$ ($\langle d_{m'}^l(\mathbf{r}_3) | d_m^u(0) \rangle$) is then related to $\langle d_{m'}^l(\mathbf{r}_1) | d_m^u(0) \rangle$ through a $2\pi/3$ ($-2\pi/3$) rotation operation $\hat{\mathcal{R}}\left(\pm\frac{2\pi}{3}\right)$ on $d_m^u(0)$ and $d_{m'}^l(\mathbf{r}_1)$, where $\hat{\mathcal{R}}\left(\pm\frac{2\pi}{3}\right) d_m^{u,l} = e^{\mp i\frac{2\pi}{3}m} d_m^{u,l}$.

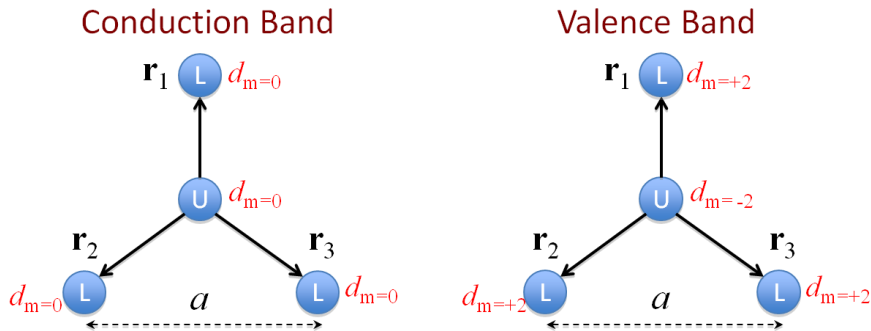

**Figure S1| Nearest neighbor interlayer hopping of metal atoms at the K point in AB-stacked bilayers.** Atoms located in the upper (lower) layer are indicated by U (L) while red labels indicate the corresponding orbitals of the K point band edge Bloch functions.

Since $m=0$, $\langle d^l_{m=0}(\mathbf{r}_1)|d^u_{m=0}(0)\rangle = \langle d^l_{m=0}(\mathbf{r}_2)|d^u_{m=0}(0)\rangle = \langle d^l_{m=0}(\mathbf{r}_3)|d^u_{m=0}(0)\rangle$ and interlayer hopping for conduction electrons at the K point is:

$$\sum_{n=1,2,3} e^{i\mathbf{K}\cdot\mathbf{r}_n}\langle d^l_{m=0}(\mathbf{r}_n)|d^u_{m=0}(0)\rangle \propto \sum_{n=1,2,3} e^{i\mathbf{K}\cdot\mathbf{r}_n} = 0.$$

For the valence band $\langle d^l_{m=+2}(\mathbf{r}_2)|d^u_{m=-2}(0)\rangle = e^{i2\pi/3}\langle d^l_{m=+2}(\mathbf{r}_1)|d^u_{m=-2}(0)\rangle$ and $\langle d^l_{m=+2}(\mathbf{r}_3)|d^u_{m=-2}(0)\rangle = e^{-i2\pi/3}\langle d^l_{m=+2}(\mathbf{r}_1)|d^u_{m=-2}(0)\rangle$. Thus interlayer hopping for holes at the K point is finite:

$$\sum_{n=1,2,3} e^{i\mathbf{K}\cdot\mathbf{r}_n}\langle d^l_{m=+2}(\mathbf{r}_n)|d^u_{m=-2}(0)\rangle = 3\langle d^l_{m=+2}(\mathbf{r}_1)|d^u_{m=-2}(0)\rangle.$$

We note that the conduction band edge wavefunctions also have a small component of $d_{xz}$ and $d_{yz}$ orbitals [S1]. At the K points in WSe$_2$, they introduce an interlayer tunneling $t_\perp \sim 0.4$ meV [S2]. However, the conduction band also has a spin splitting $2\lambda \sim 40$ meV [S1]. In AB stacked bilayer, this splitting has a valley and layer dependent sign, and corresponds to the energy cost for the interlayer hopping (see Fig. 1 in main text). Such a large ratio of $2\lambda/t_\perp$ virtually suppresses interlayer hopping of conduction electrons near K points.

**S2. Broadband monolayer and bilayer WSe$_2$ PL spectra**

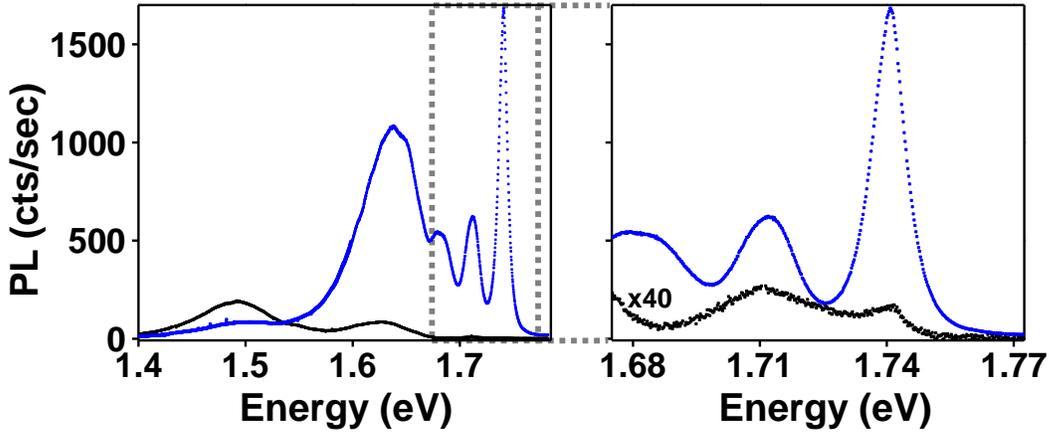

**Figure S2| Comparison of monolayer and bilayer WSe$_2$ PL energies.** Left panel: broadband PL spectrum of monolayer (blue) and bilayer (black) WSe$_2$. Right Panel: zoom-in of spectra within dashed boxed of left panel, showing bilayer WSe$_2$ data multiplied by a factor of 40. Emission at 1.74 eV for both monolayers and bilayers corresponds to X$^o$, while emission at 1.71 eV is from X$^-$.

## S3. Broadband bi-WSe$_2$ differential reflectivity spectrum

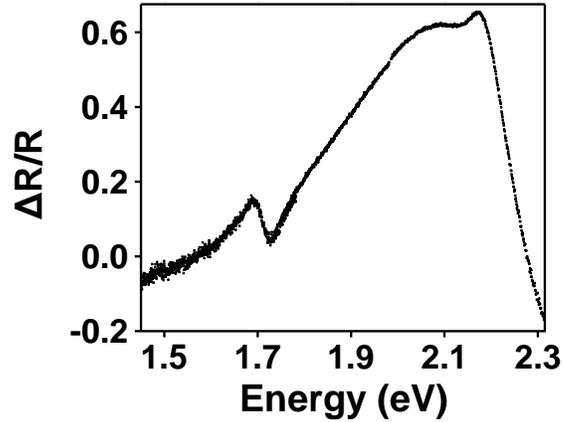

**Figure S3| Broadband bi-WSe$_2$ ΔR/R spectrum.** Broadband differential reflectivity spectrum for bi-WSe$_2$ at $V_g = 0$ showing the lowest energy absorption feature at ~1.7 eV.

## S4. Gate dependence of bi-WSe$_2$ peak fitting parameters

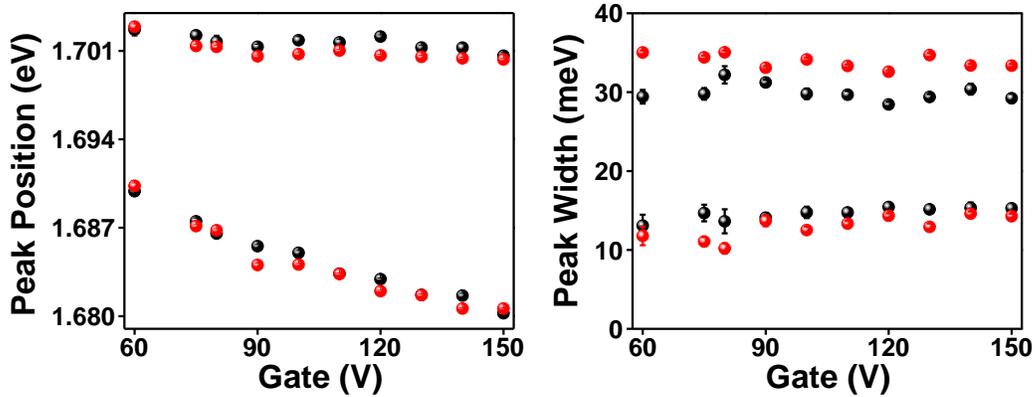

**Figure S4| Gate dependence of doublet peak fitting parameters.** Left panel: peak position as a function of gate voltage for split $X^-$ peaks. Right panel: corresponding peak widths vs. gate voltage show little variation. Upper and lower data points correspond to peaks I and II, respectively (see main text), while black (red) points correspond to fits for circularly (linearly) polarized excitation.

## S5. Second-harmonic generation in mono-, bi- and tri-layers of WSe$_2$

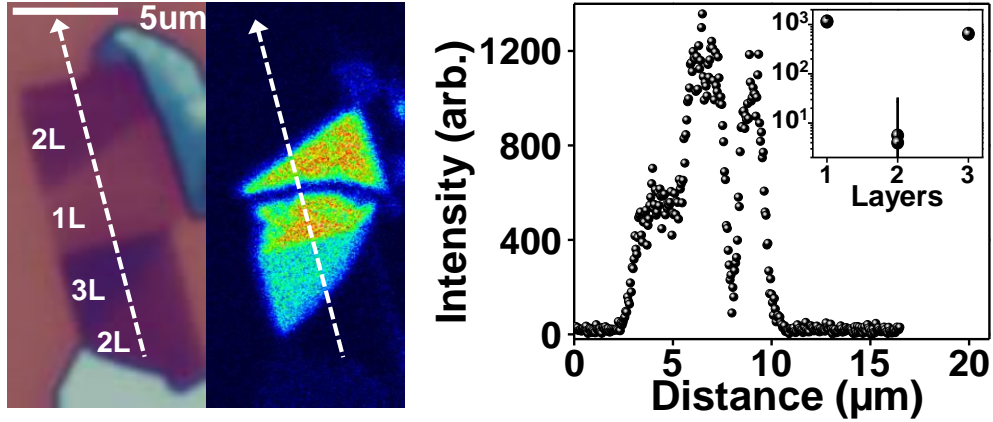

**Figure S5| Layer-dependent second-harmonic generation.** Left panel: optical microscope image of exfoliated WSe$_2$ with indicated layer number, alongside corresponding map of generated second harmonic intensity. Right panel: second harmonic intensity along line cut indicated by white dashed line in left panel. Inset: second harmonic intensity as a function of number of layers, showing contrast $>10^2$ between odd/even layers [S3-S5]. The line cut and inset plots show that the SHG signal from bilayers is within the noise of the background signal and can be ignored.

## S6. Bilayer band structure in out-of-plane E-field

Following the discussions in section S1, at the $\pm\mathbf{K}$ points, the Hamiltonian of AB stacked bilayers in an out-of-plane electric field $E$ can be written as [S6],

$$\begin{bmatrix} \Delta - \tau_z s_z \lambda_c + \frac{Ed}{2} & 0 & 0 & 0 \\ 0 & \Delta + \tau_z s_z \lambda_c - \frac{Ed}{2} & 0 & 0 \\ 0 & 0 & -\tau_z s_z \lambda_v + \frac{Ed}{2} & t_\perp \\ 0 & 0 & t_\perp & \tau_z s_z \lambda_v - \frac{Ed}{2} \end{bmatrix}.$$

The basis is $\{|d_{z^2}^u\rangle, |d_{z^2}^l\rangle, 1/\sqrt{2}(|d_{x^2-y^2}^u\rangle - i\tau_z|d_{xy}^u\rangle), 1/\sqrt{2}(|d_{x^2-y^2}^l\rangle + i\tau_z|d_{xy}^l\rangle)\}$, where the superscripts "$u$" and "$l$" denote the "upper" and "lower" layer, respectively, with interlayer separation $d$. The monolayer band gap is denoted $\Delta$, with $\lambda_c$ ($\lambda_v$) the spin-valley coupling of electrons (holes) [S1]. Interlayer hopping for holes is $t_\perp$, whereas it vanishes for electrons. $\tau_z = \pm 1$ is the valley index of bilayer bands and $s_z$ denotes the Pauli spin matrices. The out-of-plane electric field introduces an energy difference between the upper and lower layer, $\Delta_{Ec} = Ed$.

Holes at the K points then have eigenenergies $\pm\sqrt{(\tau_z s_z \lambda_v - Ed/2)^2 + |t_\perp|^2}$. Due to the large valence band spin splitting $2\lambda_v$, we focus only on the higher lying states at the band edge (the lower lying ones are hundreds of meV away). Hole states localized predominantly in the upper layer are:

$$|\downarrow\rangle'_{Kv} = \sqrt{1-|\alpha_1(E)|^2}|\downarrow\rangle_{u,Kv} + \alpha_1(E)|\downarrow\rangle_{l,Kv},$$
$$|\uparrow\rangle'_{-Kv} = \sqrt{1-|\alpha_1(E)|^2}|\uparrow\rangle_{u,-Kv} + \alpha_1(E)|\uparrow\rangle_{l,-Kv},$$

and have an eigenenergy $E_u = \sqrt{(\lambda_v + Ed/2)^2 + |t_\perp|^2}$; hole states largely localized in the lower layer are:

$$|\uparrow\rangle'_{Kv} = \sqrt{1-|\alpha_2(E)|^2}|\uparrow\rangle_{l,Kv} + \alpha_2(E)|\uparrow\rangle_{u,Kv},$$
$$|\downarrow\rangle'_{-Kv} = \sqrt{1-|\alpha_2(E)|^2}|\downarrow\rangle_{l,-Kv} + \alpha_2(E)|\downarrow\rangle_{u,-Kv},$$

and have an eigenenergy $E_l = \sqrt{(\lambda_v - Ed/2)^2 + |t_\perp|^2}$ (see Fig. S6). Here we use the shorthand notation $|\downarrow\rangle_{u,Kv} \equiv \frac{|d^u_{x^2-y^2}\rangle - i|d^u_{xy}\rangle}{\sqrt{2}} \otimes |\downarrow\rangle_K$. The coefficients $\alpha_1(E) = t_\perp / \sqrt{\left(E_u + \lambda_v + \frac{Ed}{2}\right)^2 + |t_\perp|^2}$ and $\alpha_2(E) = t_\perp / \sqrt{\left(E_l + \lambda_v - \frac{Ed}{2}\right)^2 + |t_\perp|^2}$. The electric field thus induces a spin splitting at the valence band edge at the K points $\Delta_{Ev} = E_u - E_l$, which can be approximated as $\Delta_{Ev} \approx \frac{\lambda_v}{\sqrt{\lambda_v^2 + |t_\perp|^2}} Ed$ when $Ed \ll \lambda_v$, whereas for the conduction bands, we simply have the spin splitting $\Delta_{Ec} = Ed$. Our white light differential reflectivity measurements on monolayer $WSe_2$ show $2\lambda_v = 450$ meV, in good agreement with the first principle calculations in Ref. [S6] which give $2\lambda_v = 456$ meV. The same work extracts an interlayer hopping strength of $2t_\perp = 134$ meV for the valence band edges at the K points of bilayer $WSe_2$, from which we calculate $\Delta_{Ev} \approx 0.96 \Delta_{Ec}$, and $\alpha_1(E) \approx \alpha_2(E) \approx 0.14$ for $Ed \ll \lambda_v$. The resulting difference between conduction and valence band shifts, $\Delta_{Ec} - \Delta_{Ev}$, contributes to the trion splitting observed at large gate voltages (see Fig. 3 and 4 in main text).

It should be noted that the interlayer hole hopping $t_\perp$ will induce a change in the direct band gap at the K points of bilayers from that of monolayers. At zero electric field, we have the band gap values $\Delta - \lambda_c - \lambda_v$ and $\Delta - \lambda_c - \sqrt{\lambda_v^2 + |t_\perp|^2}$ for monolayers and bilayers, respectively. In $WSe_2$, because of the large $\lambda_v/t_\perp$ ratio, we expect a small difference between the monolayer and bilayer direct band gap at K points: $\sqrt{\lambda_v^2 + |t_\perp|^2} - \lambda_v \sim 9$ meV, which is much smaller than the trion charging energy of ~30 meV. As a result, the resonances of excitons and trions formed at the K valleys of bilayers are nearly aligned with those of monolayers (Fig. S2). In contrast, in $MoSe_2$ the smaller value of $2\lambda_v = 182$ meV gives rise to a larger redshift of the exciton and trion lines in bilayers when compared to monolayers, as observed (see e.g. supplementary information of [S7]).

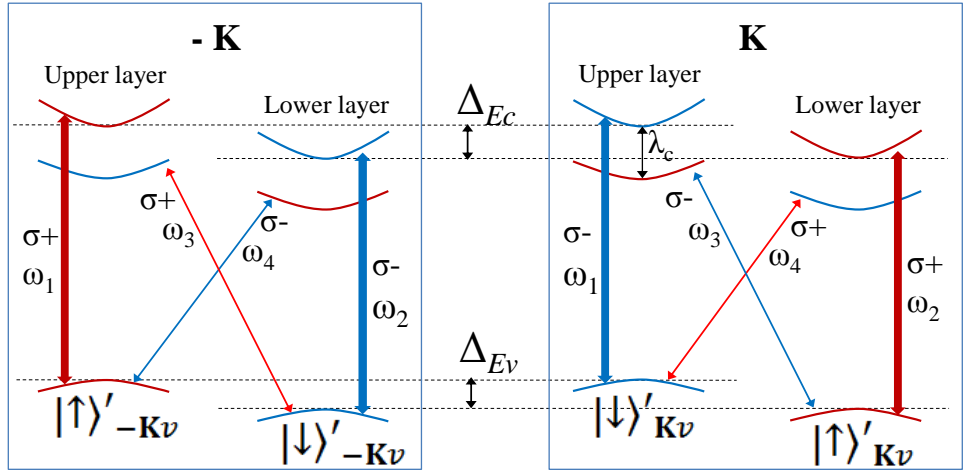

**Figure S6| Band structure of bi-WSe$_2$ under applied electric field.** The creation process of electron-hole pairs at resonances $\omega_1$, $\omega_2$, $\omega_3$ and $\omega_4$ via $\sigma^+$ ($\sigma^-$) polarized excitation is denoted by red (blue) double arrows. Arrow thickness denotes transition strength. $\Delta_{Ec}$ ($\Delta_{Ev}$) is the interlayer conduction (valence) band splitting under the effect of an out-of-plane electric field.

## S7. PL emission under circularly polarized excitation

An applied $\sigma^+$ ($\sigma^-$) polarized laser can excite electron-hole pairs through four transitions, as illustrated by red (blue) double arrows in Fig. S6. The arrow thickness denotes strength of the transition. The weak cross transitions denoted by thin arrows are due to the fact that the layer polarization of hole states near K points is ~96%, not fully polarized, a residual effect of the interlayer hopping.

Below we consider applying an incident laser with $\sigma^+$ polarization, which will excite electron-hole pairs through the four transitions denoted by the red arrows in Fig. S7. In emission, $\sigma^+$ PL will come predominantly from the two transitions marked by the thicker arrows with energies $\omega_1$ and $\omega_2$. Since the difference between $\Delta_{Ev}$ and $\Delta_{Ec}$ contributes to the difference between $\omega_1$ and $\omega_2$, in sufficiently large electric fields, this splitting may become resolvable.

The $\sigma^-$ PL signal component is a consequence of electron spin relaxation (intra- and inter-valley). We expect that intra-valley spin relaxation is more efficient than inter-valley relaxation as the latter involves a simultaneous spin and valley flip. When $\Delta_{Ev}$ and $\Delta_{Ec}$ are much larger than the temperature, the spin relaxation channel from higher energy to lower energy states by phonon emission will also be much more efficient than the backward channel requiring phonon absorption. Thus the dominant spin relaxation channel is the one-way channel depicted by the green dotted arrow in Fig. S7. Therefore, we expect $\sigma^-$ PL at resonance $\omega_2$ to be much larger than at $\omega_1$.

In summary, under $\sigma^+$ excitation we expect comparable $\sigma^+$ PL emission at both $\omega_1$ and $\omega_2$, while $\sigma^-$ PL emission resulting from carrier spin relaxation is predominantly at the lower resonance $\omega_2$. When $\omega_1 - \omega_2$ is smaller than the peak linewidth, this manifests as a red shift of the $\sigma^-$ PL.

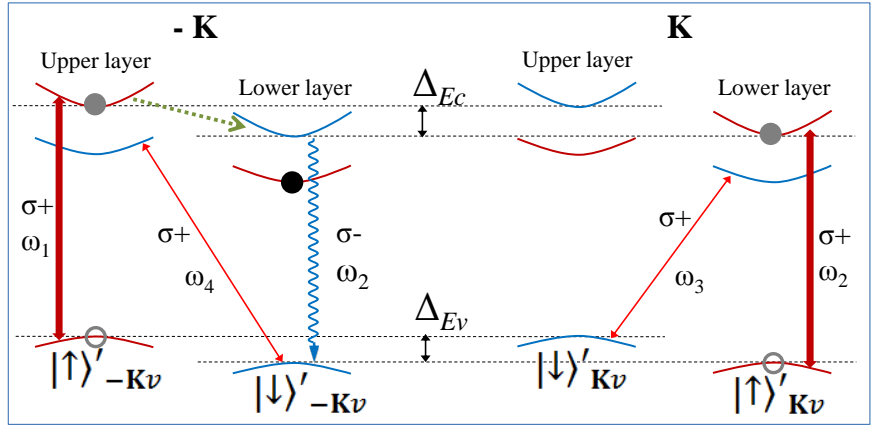

**Figure S7| Allowed excitation and emission channels for $\sigma^+$ polarized excitation.** Transitions driven by $\sigma^+$ polarized excitation are denoted by red arrows, with line thickness denoting transition strength. Gray circles denote photo-excited electron-hole pairs at either the K or -K valley. The black circle denotes the excess electron in X- trion configurations. The dominant emission channels are $\sigma^+$ polarized at frequencies $\omega_1$ and $\omega_2$. Emission of $\sigma^-$ PL at $\omega_2$ is indicated by the wavy blue arrow which becomes possible following the intra-valley spin flip of the photo-excited electron accompanied by emission of energy into the environment (green dashed arrow). The spin flipped electron can then recombine with holes photo-excited through the weak cross transition ($\omega_4$) and emit a $\sigma^-$ photon.

### S8. X⁻ peak splitting

In the experiment a doublet peak feature is observed in the X⁻ spectrum for large gate voltages (main text Figs. 3a and 4a). We attribute this to the two different X⁻ configurations shown in Fig. S8a, where the X⁻ energy can be obtained from the X⁰ resonance by subtracting the trion charging energy. In the $\omega_1$ configuration (top panel), the excess electron and the neutral exciton are in different layers (interlayer trion) and we write its charging energy as $E_{c1}$. In contrast, for the $\omega_2$ configuration (bottom panel), the excess electron and neutral exciton are in the same layer (intralayer trion) and we write its charging energy as $E_{c2}$. Because of the layer separation, $E_{c1} < E_{c2}$ (see Fig. S8a). Thus, in the X⁻ spectrum the peak splitting is $\omega_1 - \omega_2 = (\Delta_{Ec} - \Delta_{Ev}) + (E_{c2} - E_{c1})$. Since the layer separation (7 Å) is comparable to the exciton Bohr radius ($a_B \sim 1$ nm) [S8], and the monolayer trion charging energy is $E_{c2} \approx 30$ meV, we expect the charging energy difference $E_{c2} - E_{c1}$ can be as large as several meV.

In contrast, for the X⁺ spectrum, the $\omega_1$ configuration corresponds to *intralayer* trions with the excess hole and neutral exciton in the same layer (top panel Fig. S8b), and a charging energy $E_{c1}$. For the $\omega_2$ configuration we have interlayer trions with the excess hole and neutral exciton in different layers (bottom panel Fig. S8b), and a charging energy $E_{c2}$. Now $E_{c1} > E_{c2}$, and the peak splitting is $\omega_1 - \omega_2 = (\Delta_{Ec} - \Delta_{Ev}) -$ ($E_{c1} - E_{c2}$), where the two contributions largely cancel. Consequently, the peak splitting $|\omega_1 - \omega_2|$ is larger for X⁻ than X⁺ and thus more readily observed.

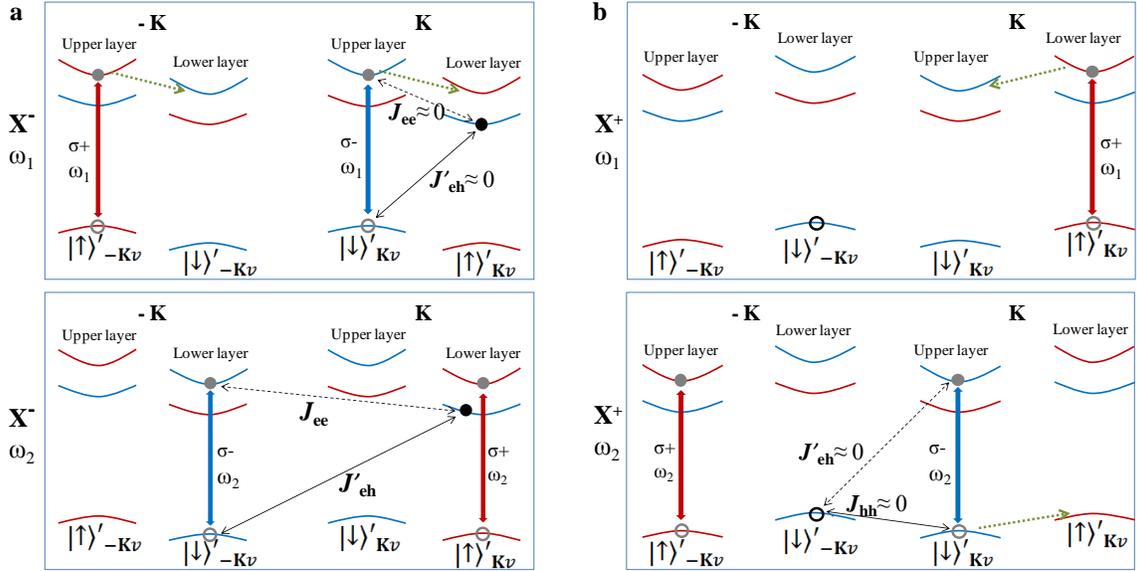

**Figure S8| $X^-$ and $X^+$ configurations under linearly polarized excitation. a,** $X^-$ configurations with emission energy $\omega_1$ (upper panel) and $\omega_2$ (lower panel), where an electron-hole pair (gray dot and circle) in a coherent superposition in the two valleys can lead to linearly polarized emission. Black dot denotes the excess electron in the lowest energy band (lower layer). Dashed green arrows represent spin relaxation processes and black double arrows indicate exchange interactions. Intervalley coherence at $\omega_1$ (interlayer trions) is less affected by the excess electron than at $\omega_2$ (intralayer trions) since the electron is on the layer opposite the recombining electron-hole pair. **b,** For $X^+$, the excess hole is in the lowest energy band (the lower layer). A coherent superposition in ±K valleys is not possible for intralayer positive trions at $\omega_1$ (top panel) since the excess hole comprising the trion resides within the same layer, as in monolayers [S9]. For interlayer positive trions where the electron-hole pair is in the upper layer, opposite the excess hole (bottom panel), a valley superposition is allowed and experiences negligible exchange interactions.

### S9. PL emission for linearly polarized excitation

A linearly polarized incident laser will create electron-hole pairs in a coherent superposition of states at K and –K. If the intervalley coherence can be preserved until electron-hole recombination, linearly polarized PL is emitted. Bilayers of transition metal dichalcogenides are qualitatively different from monolayers because of the additional layer degree of freedom. Below we analyze how valley coherence of a photo-excited electron-hole pair can be affected by the excess electron or hole in the X⁻ or X⁺ configuration in bilayers. We consider the condition where the electric field induced Zeeman splitting between $\Delta_{Ev}$ and $\Delta_{Ec}$ is large, such that the excess electron (hole) is in the lower energy layer, as denoted by the black filled (open) circle in the lowest (highest) conduction (valence) band in Fig. S8a (Fig. S8b).

X⁻ emission occurs at either energy $\omega_1$ or $\omega_2$. We first look at the emission at $\omega_2$ which comes from the intralayer trion configuration shown in lower panel of Fig. S8a, i.e. the photo-excited electron-hole pair (gray circles) is in same layer as the excess electron. Non-zero exchange interactions with the excess electron can destroy the intervalley coherence of the electron-hole pair, just as in the monolayer case [S9]. Hence, linear polarization is largely suppressed at this energy and one expects both co-polarized (i.e. identical to the excitation polarization) and cross-polarized (i.e. orthogonal to the excitation polarization) PL emission from intralayer trions. We now turn to X⁻ emission at $\omega_1$, which comes from the interlayer trion configuration depicted in the upper panel of Fig. S8a. With the photo-excited electron-hole pair in the layer opposite the excess electron, the exchange interaction is substantially suppressed, such that intervalley coherence can be preserved and large linearly polarized PL can be observed at $\omega_1$. This is in good agreement with the observed linear polarization behavior of the X⁻ doublet with applied gate.

In addition to the above exchange interactions which suppress linear polarization at $\omega_2$, carrier spin relaxation as discussed in the previous section (green dotted arrows in Fig. S8a) will contribute unpolarized emission at $\omega_2$ as well, further suppressing linearly polarized PL at this emission energy.

For the X⁺ trion the situation is different. Emission at $\omega_1$ is from the intralayer trion configuration shown in upper panel of Fig. S8b, where the photo-excited electron-hole pair is in the same layer as the excess hole. Similar to the monolayer case, because of the Pauli exclusion principle, X⁺ PL cannot exhibit linear polarization at $\omega_1$ [S9]. Emission at $\omega_2$ is from the interlayer trion configuration shown in the lower panel of Fig. S8b, where the photo-excited electron-hole pair is in the layer opposite the excess hole. Linearly polarized emission is thus allowed at $\omega_2$. As before, interlayer exchange interactions are strongly suppressed so that at this emission energy, the major depolarizing mechanism is the spin-relaxation of carriers denoted by the dotted green arrow.

## S10. Supplementary References